\documentstyle[twoside,12pt]{article}
\normalsize
\def\greaterthansquiggle{\raise.3ex\hbox{$>$\kern-.75em\lower1ex\hbox{$\sim$}}}
\def\lessthansquiggle{\raise.3ex\hbox{$<$\kern-.75em\lower1ex\hbox{$\sim$}}}
\newcommand{\bdi}{\begin{displaymath}}
\newcommand{\edi}{\end{displaymath}}
\newcommand{\bfi}{\begin{figure}}
\newcommand{\efi}{\end{figure}}

\newcommand{\beq}{\begin{equation}}
\newcommand{\eeq}{\end{equation}}

\newcommand{\gaf}{\gamma_{5}}

\newcommand{\beqa}{\begin{eqnarray}}
\newcommand{\eeqa}{\end{eqnarray}}
\newcommand{\no}{\nonumber}

\newcommand{\ra}{\rightarrow}

\newcommand{\wt}{\widetilde}

\def\au{{\setbox0=\hbox{\lower1.36775ex%
\hbox{''}\kern-.05em}\dp0=.36775ex\hskip0pt\box0}}
\def\ao{{}\kern-.10em\hbox{``}}

\begin{document}
\bibliographystyle{plain}

\begin{titlepage}
\begin{flushleft} 
FSUJ TPI 12/96
\end{flushleft}
\begin{flushright}
September, 1996
\end{flushright}

\vspace{1cm}
\begin{center}
{\Large \bf Decay widths in the massive Schwinger model}\\[1cm]
C. Adam* \\
Friedrich-Schiller-Universit\"at Jena, Theoretisch-Physikalisches Institut \\
Max-Wien Platz 1, D-07743 Jena, Germany
\vfill
{\bf Abstract} \\
\end{center}
By a closer inspection of the massive Schwinger model within mass perturbation
theory we find that, in addition to the $n$-boson bound states, a further type
of hybrid bound states has to be included into the model. Further we explicitly
compute the decay widths of the three-boson bound state and of the lightest
hybrid bound state.

\vfill

$^*)${\footnotesize permanent address: Inst. f. theoret. Physik d. Uni Wien \\
Boltzmanngasse 5, 1090 Wien, Austria \\
email address: adam@pap.univie.ac.at}
\end{titlepage}

\section{Introduction}
The massive Schwinger model is two-dimensional QED with one massive fermion.
In this model there are instanton-like gauge field configurations present, and,
therefore, a $\theta$ vacuum has to be introduced as a new, physical vacuum
(\cite{CJS,Co1}). Further, confinement is realized in this model in the sense
that there are no fermions in the physical spectrum (\cite{AAR,GKMS}). The
fermions form charge neutral bosons, and only the latter ones exist as physical
particles. 
The fundamental particle of the theory is a massive, interacting boson with
mass $\mu =M_1$ (Schwinger boson). 
In addition, there exist $n$-boson bound states. The two-boson
bound state is stable (mass $M_2$), whereas the higher bound states may decay
into $M_1$ and $M_2$ particles (\cite{Co1}). 
All these features have been discussed in
\cite{GBOUND} within mass perturbation theory (\cite{Co1}, 
\cite{FS1} -- \cite{SMASS}), 
which uses the exacly soluble massless Schwinger model (\cite{AAR}, 
\cite{Sc1} -- \cite{Adam}) as a starting point.

Here we will find that another type of unstable bound states has to be included
into the theory, namely hybrid bound states composed of $M_1$ and $M_2$
particles. In addition, we will compute the decay widths of the $M_3$ bound
state and of the lightest hybrid bound state (which consists of one $M_1$ and
one $M_2$ and has mass $M_{1,1}$).

\section{Bound states}

For later convenience we define the functions
\beq
E_\pm (x)=e^{\pm 4\pi D_\mu (x)} -1
\eeq
and their Fourier transforms $\wt E_\pm (p)$, where $D_\mu (x)$ is the massive
scalar propagator. As was discussed in \cite{GBOUND}, all the $n$-boson bound
state masses $M_n$ may be inferred from the two-point function ($P=\bar\Psi
\gaf \Psi$, $S=\bar\Psi \Psi$, $S_\pm =\bar\Psi \frac{1}{2}(1\pm \gaf)\Psi $)
\beq
\Pi (x):=\delta (x)+g\langle P(x)P(0)\rangle
\eeq
(or $g\langle S(x)S(0)\rangle$ for even bound states) where $g=m\Sigma +o(m^2)$
is the coupling constant of the mass perturbation theory for vanishing vacuum
angle $\theta=0$ (the general $\theta$ case we discuss in a moment); $\Sigma$
is the fermion condensate of the massless model. $\Pi (x)$ is related to the
bosonic $n$-point functions of the theory via the Dyson-Schwinger equations 
\cite{GBOUND}. In momentum space $\wt \Pi (p)$ may be resummed,
\beq
\wt \Pi (p)=\frac{1}{1-g\wt{\langle PP\rangle}_{\rm n.f.}(p)}
\eeq
where n.f. means non-factorizable and denotes all Feynman graphs that may not
be factorized in momentum space. In lowest order $\wt{\langle PP\rangle}_{\rm
n.f.}$ is
\beq
\wt{\langle PP\rangle}_{\rm n.f.} (p)=\frac{1}{2}(\wt E_+ (p)-\wt E_- (p))
\eeq
(and with a + for $\wt{\langle SS\rangle}_{\rm n.f.}$). Expanding the
exponential one finds $1-g\sum_{n=1}^\infty \frac{(4\pi)^n}{n!}
\wt{D_\mu^n}(p)$ in the denominator of (3) (more precisely, the odd powers for
$\wt{\langle PP\rangle}_{\rm n.f.}$, the even powers for $\wt{\langle SS
\rangle}_{\rm n.f.}$). At $p^2 =(n\mu)^2$,
$\wt{D_\mu^n}(p)$ is singular, therefore there are mass poles $p^2 =M_n^2$ 
slightly below the $n$-boson thresholds. Further $\wt{D_\mu^m}(p)$ have
imaginary parts at $p^2 =M_n^2$ for $m<n$, therefore decays into $mM_1$ are
possible (more precisely, for the parity conserving case $\theta =0$, only
odd $\ra$ odd or even $\ra$ even decays are possible).

Up to now we did not mention the $M_2$ particle, although decays into some
$M_2$ are perfectly possible. So where is it?
The boson bound states are found by a resummation, so a further resummation is
a reasonable idea. Let us look at the $M_1 +M_2$ final state for definiteness. 
Within $g\wt{\langle PP\rangle}_{\rm n.f.}(p)$ we may find the following term
\beq
H(p):= \int\frac{d^2 q}{(2\pi)^2}g\wt{\langle PP\rangle}_{\rm n.f.}(q)
\wt \Pi (q) g\wt{\langle PP\rangle}_{\rm n.f.}(q) 4\pi\wt D_\mu (p-q).
\eeq
It is simply a loop where we have selected one boson to run along the first
line, whereas the $g\wt{\langle PP\rangle}_{\rm n.f.}(q)$ 
and $\wt\Pi (q)$ run along
the other line. $H(p)$ is a loop, therefore it is non-factorizable.
The additional $(g\wt{\langle PP\rangle}_{\rm n.f.}(q))^2$ factor is necessary,
because $\wt\Pi (q)$ starts at zeroth order ($\wt\Pi (q) =1+g\wt{\langle
PP\rangle}_{\rm n.f.}(q)+\ldots$), and without this factor we would 
include some
diagrams into $H(p)$ that were already used for the $n$-boson bound-state
formation (double counting).

The claim is that $H(p)$ has a threshold singularity precisely at $p^2 =(M_1
+M_2)^2$, and therefore an imaginary part for $p^2 >(M_1 +M_2)^2$. But this is
easy to see. At $q^2 =M_2^2$ $\wt\Pi(q)$ has the $M_2$ one-particle singularity
and $g\wt{\langle PP\rangle}_{\rm n.f.}(M_2)\equiv 1$. Therefore, near $p^2
=(M_1 +M_2)^2$, $H(p)$ is just the $M_1 ,M_2$- two-boson loop (up to a
normalization constant). 

Observe that this line of reasoning is not true for higher bound states, $p^2
\simeq (M_n +M_1)^2 ,n>2$. $\wt\Pi (M_n)$ contains imaginary parts and is not
singular for $n>2$ (because the $M_n$ are unstable), and therefore $H(p)$ has
no thresholds at higher $p^2$.

A further consequence is that $H(p)$ gives rise to a further mass pole slightly
below $p^2 =(M_1 +M_2)^2$ in (3).

These considerations may be generalized, and we find $n_1 M_1 +n_2 M_2$
particle-production thresholds at $p^2 =(n_1 M_1 +n_2 M_2)^2$ and (unstable)
$n_1 M_1 +n_2 M_2$-bound states slightly below.

After all, this is not so surprizing. The $M_2$ are stable particles and
interacting via an attractive force. In two dimensions this {\em must} give
rise to a bound state formation. (Similar conclusions may be drawn from
unitarity when $M_2$-scattering is considered, \cite{SCAT}.)

Before starting the actual computations, we should generalize to arbitrary
$\theta \ne 0$. There the coupling constant is complex, $g\ra g_\theta
,g_\theta^*$, and, because of parity violation, the Feynman rules acquire a
matrix structure (the propagators are $2\times 2$ matrices, the vertices
tensors, etc.). The exact propagator may be inverted, analogously to (3), and
leads to (see \cite{GBOUND})
\beq
\frac{{\cal M}_{ij}}{1-\alpha -\alpha^* +\alpha \alpha^* -\beta \beta^* }
\eeq
\beq
\alpha (p)=g_\theta \wt{\langle S_+ S_+ \rangle}_{\rm n.f.}(p) \quad ,\quad
\beta (p)=g_\theta \wt{\langle S_+ S_- \rangle}_{\rm n.f.}(p)
\eeq
and ${\cal M}_{ij}$ ($i,j=+,-$) gives the $\wt{\langle S_i S_j \rangle}$
component of the propagator. For our considerations only the denominator in
(6) is important. In leading order
\beq
\alpha (p) =g_\theta \wt E_+ (p)\quad ,\quad \beta (p)=g_\theta \wt E_- (p)
\quad ,\quad g_\theta =\frac{m\Sigma}{2} e^{i\theta}
\eeq
and the denominator reads
\beq
1-m\Sigma\cos\theta \wt E_+ (p) +\frac{m^2 \Sigma^2}{4}(\wt E_+^2 (p)-\wt E_-^2
(p)).
\eeq
Inserting the $n$-boson functions ($d_n
(p):=\frac{(4\pi)^n}{n!}\wt{D_\mu^n}(p)$) results in
\beq
1-m\Sigma\cos\theta (d_1 +d_2 +\ldots )+m^2 \Sigma^2 
\Bigl( d_1 (d_2 +d_4 +\ldots) +d_3 (d_2 +d_4 +\ldots)+\ldots \Bigr)
\eeq
Now suppose we are e.g. at the $M_3$ bound state mass. Then the real part of
(10) vanishes by definition and $m\Sigma\cos\theta d_3 (M_3) =1+o(m)$, and we
get
\beq
-im\Sigma\cos\theta {\rm Im\,} d_2 (M_3) +im^2\Sigma^2 d_3 (M_3){\rm Im\,} d_2 (M_3)=
-im\Sigma (\cos\theta -\frac{1}{\cos\theta}){\rm Im\,} d_2 (M_3).
\eeq
This computation may be generalized easily, and we find that each parity
allowed decay acquires a $\cos\theta$, whereas a parity forbidden decay
acquires a $(\cos\theta - \frac{1}{\cos\theta})$ factor.

To include the decays into $M_2$ we have to perform a further resummation
analogous to above, however, the resummed contributions enter into the
functions $\alpha$, $\beta$ in a way that is perfectly consistent with our
parity considerations (a $n_1 M_1 +n_2 M_2$-state has parity $P=(-1)^{n_1}$).

\section{Bound state masses}

We are now prepared for explicit computations, but before computing decay
widths we need the masses and residues of the propagator at the various mass
poles. The masses $M_1 ,M_2 ,M_3$ have already been computed (\cite{GBOUND};
there is, however, a numerical error in the $M_2$ mass formula in
\cite{GBOUND}),
\beq
M_1^2 \equiv \mu^2 =\mu_0^2 +\Delta_1  +o(m^2) \quad ,\quad \Delta_1 =4\pi
m\Sigma\cos\theta
\eeq
\beq
M_2^2 =4\mu^2 -\Delta_2 \quad ,\quad \Delta_2 =\frac{4\pi^4 m^2 \Sigma^2 \cos^2
\theta}{\mu^2}
\eeq
\beq
M_3^2 =9\mu^2 -\Delta_3 \quad ,\quad \Delta_3 \simeq 
6.993 \mu^2 \exp (-0.263\frac{\mu^2}{m\Sigma \cos\theta})
\eeq
and the three-boson binding energy is smaller than polynomial in the coupling
constant $m$ (or $g$). 

In leading order the $n$-th mass pole is the zero of the function
\beq
f_n (p^2)=1-m\Sigma\cos\theta d_n (p^2),
\eeq
therefore the residue may be inferred from the first 
Taylor coefficient around $(p^2 -M_n^2 )$,
\beq
f_n (p^2) \simeq c_n (p^2 -M_n^2).
\eeq
The $c_n$ may be inferred from the computation of the mass poles
(\cite{GBOUND}) and are
related to the binding energies. Explicitly they read
\beq
c_1 =\frac{1}{4\pi m\Sigma\cos\theta}=\frac{1}{\Delta_1}
\eeq
\beq
c_2 =\frac{\mu^2}{8\pi^4 (m\Sigma\cos\theta)^2}=\frac{1}{2\Delta_2}
\eeq
\beq
c_3 =\frac{m\Sigma\cos\theta}{0.263 \mu^2 \Delta_3}
\eeq
The mass $M_{1,1}$ is the solution of $1=(g_\theta +g_\theta^* )H(p)$, which
looks difficult to solve. However, there is an approximation. At threshold 
$\wt \Pi
(q)$ equals the $M_2$ propagator, so this may be a reasonable approximation
provided that the binding energy is sufficiently small, $\Delta_{1,1}\equiv
(M_1 +M_2)^2 -M_{1,1}^2 <\Delta_2$. In this approximation we have for $M_{1,1}$
\beqa
1 &=& m\Sigma\cos\theta\int\frac{d^2 q}{(2\pi)^2}\frac{8\pi^4
m\Sigma\cos\theta}{\mu^2 (q^2 -M_2^2)}\frac{4\pi}{(p-q)^2 -M_1^2} \no \\
&=& \frac{32\pi^5 m^2 \Sigma^2 \cos^2 \theta}{2\pi \mu^2\bar w(p^2 ,M_2^2
,M_1^2 )}\Bigl( \pi +\no \\
&& \arctan \frac{2p^2}{\bar w(p^2 ,M_2^2 ,M_1^2 ) 
-\frac{1}{\bar w(p^2 ,M_2^2 ,M_1^2 )}(p^2 +M_1^2 -M_2^2)(p^2 -M_1^2 +M_2^2 )}
\Bigr)
\eeqa
\beq
\bar w(x,y,z):=(-x^2 -y^2 -z^2 +2xy+2xz+2yz)^{\frac{1}{2}}
\eeq
where we inserted the residues that may be derived from the Taylor coefficients
$c_1 ,c_2$ (17,18) (${\rm Res}_i =\frac{1}{c_i m\Sigma\cos\theta}$). 
The solution is
\beq
M_{1,1}^2 =(M_1 +M_2)^2 -\Delta_{1,1}\quad ,\quad \Delta_{1,1}=
\frac{32\pi^{10}(m\Sigma\cos\theta)^4}{\mu^6}
\eeq
which shows that our approximation is justified for sufficiently small $m$.

$M_{1,1}$ was computed in a way analogous to $M_2$ (see \cite{GBOUND}),
therefore it leads to an analogous Taylor coefficient
\beq
c_{1,1} =\frac{1}{2\Delta_{1,1}}=\frac{\mu^6}{64\pi^{10}(m\Sigma\cos\theta)^4}.
\eeq

\section{Decay width computation}

The decay widths may be inferred in a simple way from the imaginary parts of
the propagator. Generally
\beq
G(p)\sim \frac{{\rm const.}}{p^2 -M^2 -i\Gamma M}
\eeq
and $\Gamma$ is the decay width. In our case the poles have their Taylor
coefficients,
\beq
\wt \Pi (p) \sim \frac{{\rm const.}}{c_i (p^2 -M_i^2) -i{\rm Im\,} (\cdots)}
\sim \frac{{\rm const'.}}{p^2 -M_i^2 -i\frac{{\rm Im\,} (\cdots)}{c_i}}
\eeq
and therefore
\beq
\Gamma_i \sim \frac{{\rm Im\,} (\cdots)}{c_i M_i}.
\eeq
Before performing the explicit computations let us add a short remark.
The $c_i$ are related to the binding energies, $c_i \sim \frac{1}{\Delta_i}$.
Therefore, all the decay widths are restricted by the binding energies,
$\Gamma_i\sim \Delta_i$. But this is a very reasonable result. The denominator
of the propagator (25) has zero real part at $M_i^2$ and infinite real part
at the real particle production threshold. Suppose $\wt \Pi (p)$ contributes to
a scattering process (to be discussed in detail in a further publication,
\cite{SCAT}). It will give rise to a local maximum (resonance) at $p^2 =M_i^2$,
and to a local minimum at the production threshold $p^2 =M_i^2 +\Delta_i$.
Therefore the resonance width (decay width) {\em must} be bounded by
$\Delta_i$.

Now let us perform the explicit calculations. At $M_{1,1}^2$ the propagator is
\beq
\wt \Pi (p)\sim\frac{1}{c_{1,1} (p^2 -M_{1,1}^2) -im\Sigma (\cos\theta
-\frac{1}{\cos\theta}) {\rm Im\,} d_2 (p)}
\eeq
\bdi
{\rm Im\,} d_2 (p)=\frac{8\pi^2}{2w(p^2 ,M_1^2 ,M_1^2 )}
\edi
\beq
w(x,y,z)=(x^2 +y^2 +z^2 -2xy-2xz-2yz)^{\frac{1}{2}}
\eeq
leading to the decay width ($M_1 \equiv \mu$)
\beq
\Gamma_{M_{1,1}}=\frac{2^8 \pi^{12} (m\Sigma\cos\theta)^5}{9\sqrt{5}\mu^9}
(\frac{1}{\cos^2 \theta} -1) \simeq 21340 \mu(\frac{m\cos\theta}{\mu})^5
(\frac{1}{\cos^2 \theta} -1)
\eeq
($\Sigma =\frac{e^\gamma \mu}{2\pi}=0.283 \mu$)
for the decay $M_{1,1}\ra 2M_1$. This decay is parity forbidden, and therefore
$M_{1,1}$ is stable for $\theta =0$.

For the $M_3$ decay there exist two channels, $M_3 \ra M_2 +M_1 ,M_3 \ra 2M_1$,
\beq
\wt \Pi (p)\sim\frac{1}{c_3 (p^2 -M_3^2) -im\Sigma (\cos\theta
-\frac{1}{\cos\theta})\frac{4\pi^2}{w(p^2 ,M_1^2 ,M_1^2 )}
-i(m\Sigma\cos\theta)^2\frac{16\pi^5}{\mu^2 w(p^2 ,M_2^2 ,M_1^2)}}
\eeq
leading to the partial decay widths
\beq
\Gamma_{M_3 \ra 2M_1}=0.263 \frac{4\pi^2
\Delta_3}{9\sqrt{5}\mu}(\frac{1}{\cos^2 \theta}-1)
\simeq 3.608 \mu (\frac{1}{\cos^2 \theta}-1)
\exp (-0.929\frac{\mu}{m\cos\theta})
\eeq
\beq
\Gamma_{M_3 \ra M_2 +M_1}=0.263\frac{4\pi^3 \Delta_3}{3\sqrt{3}\mu}
\simeq 43.9 \mu \exp (-0.929\frac{\mu}{m\cos\theta})
\eeq
and to the ratio
\beq
\frac{\Gamma_{M_3 \ra 2M_1}}{\Gamma_{M_3 \ra M_2 +M_1}}=\frac{\frac{1}{\cos^2
\theta}-1}{\sqrt{15}\pi}.
\eeq
The latter is independent of the approximations that were used in the
computation of $M_3$ and $c_3$. Observe that $\Gamma_{M_3 \ra M_2 +M_1}$ is
larger than $\Gamma_{M_3 \ra 2M_1}$, although $M_1 +M_2 \sim M_3$. This is so
because the phase space "volume" does not rise with increasing momentum in
$d=1+1$.

Remark: there seems to be a cheating concerning the sign of $\Gamma_{M_3 \ra
2M_1}$ (see (30), (31)). This is a remnant of the Euclidean conventions that
are implizit in our computations (see e.g. \cite{GBOUND}). There the
conventions are such that $\theta$ is imaginary and, consequently, $\cos\theta
-\frac{1}{\cos\theta}\ge 0$. In a really Minkowskian computation, roughly
speaking, the roles of $E_+$ and $E_-$ in (6) are exchanged, leading to a
relative sign between odd and even states. The final results (29), (31) and
(32) are expressed for Minkowski space and for real $\theta$
($\frac{1}{\cos^2 \theta} -1\ge 0$), which explains the sign.

\section{Summary}

By a closer inspection of the massive Schwinger model we have found that its
spectrum is richer than expected earlier. In addition to the $n$-boson bound
states there exist hybrid bound states that are composed of fundamental bosons
and stable two-boson bound states. A posteriori their existence is not too
surprizing and may be traced back to the fact that particles
that attract each other form at
least one bound state in $d=2$; or it may be
understood by some unitarity arguments. For
the special case of vanishing vacuum angle, $\theta =0$, the lowest of these
hybrid bound states is even stable and must be added to the physical particles
of the theory.

Further we computed the decay widths of some unstable bound states 
and found that our results are consistent with an interpretation of the
bound states as resonances. Even more insight
into these features would be possible by a discussion of scattering, which will
be done in a forthcoming publication (\cite{SCAT}).

Of course, it would be interesting to compare our results to other approaches,
like e.g. lattice calculations.

\section*{Acknowledgement}

The author thanks the members of the Institute of Theoretical Physics of the
Friedrich-Schiller-Universit\"at Jena, where this work was done, for their
hospitality. Further thanks are due to Jan Pawlowski for helpful discussions.

This work was supported by a research stipendium of the Vienna University.

\end{document}